\documentclass[letterpaper]{article} 
\usepackage{aaai25}  
\usepackage{times}  
\usepackage{helvet}  
\usepackage{courier}  
\usepackage[hyphens]{url}  
\usepackage{graphicx} 
\usepackage{natbib}  
\usepackage{caption} 
\usepackage{booktabs} 
\usepackage{subcaption}
\usepackage{amsmath}
\usepackage{algorithm}
\usepackage{algorithmic}

\usepackage{newfloat}
\usepackage{listings}
\DeclareCaptionStyle{ruled}{labelfont=normalfont,labelsep=colon,strut=off} 
\floatstyle{ruled}
\newfloat{listing}{tb}{lst}{}
\floatname{listing}{Listing}
\title{A Framework for Evaluating Vision-Language Model Safety:\\ Building Trust in AI for Public Sector Applications}

\author{
    Maisha Binte Rashid, Pablo Rivas
}
\affiliations{
    Department of Computer Science\\School of Engineering and Computer Science\\

    Baylor University, Texas USA\\
    \{Maisha\_Rashid1, Pablo\_Rivas\}@Baylor.edu
}

\usepackage{bibentry}

\begin{document}

\maketitle

\begin{abstract}
Vision-Language Models (VLMs) are increasingly deployed in public sector missions, necessitating robust evaluation of their safety and vulnerability to adversarial attacks. This paper introduces a novel framework to quantify adversarial risks in VLMs. We analyze model performance under Gaussian, salt-and-pepper, and uniform noise, identifying misclassification thresholds and deriving composite noise patches and saliency patterns that highlight vulnerable regions. These patterns are compared against the Fast Gradient Sign Method (FGSM) to assess their adversarial effectiveness. We propose a new Vulnerability Score that combines the impact of random noise and adversarial attacks, providing a comprehensive metric for evaluating model robustness. 
\end{abstract}

%

\section{Application Context}

The framework we present addresses the pressing need for robust evaluation metrics to ensure the safety and reliability of VLMs in public missions. Our work specifically targets applications where VLMs play a critical role, such as disaster response, medical diagnostics, infrastructure management, and public education. These domains demand systems capable of maintaining performance even under adverse conditions, including adversarial manipulations.

Our proposed Vulnerability Score metric evaluates the resilience of VLMs against both random noise and targeted adversarial attacks. This makes it particularly suitable for government and non-governmental organizations tasked with deploying AI systems in high-stakes environments. For example, in disaster response scenarios, where real-time and accurate decision-making is essential, identifying and mitigating adversarial risks ensures that VLMs can be trusted to process critical visual and textual information reliably. Similarly, in medical imaging or urban planning, robustness against adversarial threats protects against errors that could have significant societal consequences.

By combining comprehensive noise analysis and adversarial impact assessment, this approach aligns with the public sector’s emphasis on trustworthy AI. While the method’s computational intensity might pose challenges for resource-constrained settings, its ability to quantify vulnerabilities and provide actionable insights outweighs these limitations in mission-critical scenarios. Consequently, the framework equips stakeholders with a powerful tool to make informed decisions about deploying and governing AI technologies in public service contexts.

\section{Introduction}
With the rapid advancement of artificial intelligence, VLMs have become deeply embedded in our daily lives. As of 2023, over six billion people worldwide use smartphones~\cite{10.1177/21582440231219538}, many of which are enabled with VLM-based features like augmented reality, real-time translation, and intelligent personal assistants. The increasing integration of artificial intelligence (AI) into a wide range of public sector missions follows from disaster response and medical diagnostics to infrastructure management and educational initiatives, there is a growing imperative to ensure that such systems are both reliable and safe. The integration of VLMs into governmental and non-governmental services demands high standards of trust, fairness, and resilience, especially when these models operate under challenging conditions or encounter adversarial manipulations.

An adversarial attack involves manipulating a machine learning model by using carefully crafted data to exploit its vulnerabilities, potentially leading to incorrect predictions. While there has been extensive research on adversarial attacks in computer vision and natural language processing, \citet{binte2024navigating} identified a research gap in effectively measuring adversarial attacks and conducting safety assessments for VLMs. Existing studies have primarily focused on developing attack methodologies and their defense mechanisms, with little attention given to reliable measurement of risk assessment and developing evaluation metrics to assess vulnerability. 

To ensure safety and trustworthiness in public service AI systems, we must develop comprehensive and standardized metrics that reliably quantify risk, evaluate robustness, and identify vulnerabilities. Public missions often deal with different kinds of stakeholders, diverse data sources, and time constraints. It becomes not only a technical issue but also a public policy necessity to guarantee AI's consistent performance while maintaining equity, transparency, and social well-being.

This paper introduces a novel approach to efficiently quantifying and evaluating the adversarial vulnerabilities of VLMs, with a specific focus on the CLIP model~\cite{radford2021learning}. Our approach combines a comparison to well-known adversarial attack approaches with an analysis of model performance under different noise conditions. Our goal is to present a thorough evaluation of VLM vulnerability by looking at the model's response to Gaussian, Salt and Pepper, and Uniform noise. Our research contributes to the field in several ways:
\begin{itemize}
    \item We conduct an experiment by applying incremental noise levels to assess the threshold at which misclassification occurs in image classification task.
    \item We compare the effectiveness of our noise-based perturbations against the FGSM, a well-established adversarial attack technique.
    \item We propose a new evaluation metric that calculates a Vulnerability Score by combining the impact of both random noise and targeted adversarial attacks. 
\end{itemize}

In situations where public trust and mission success are crucial, we hope that these measurements can help policymakers, developers, and community organizations make better decisions concerning the deployment and regulation of AI systems. 
\section{Background Study}
Adversarial attacks on neural networks often involve crafting input perturbations to cause model misclassifications. One of the first and best known attack is the FGSM, which adds noise aligned with the gradient of the loss to the input data~\cite{goodfellow2014explaining}. Building on this, Projected Gradient Descent (PGD) applies iterative gradient-based perturbations, projecting the adversarial example back onto the valid input space at each step~\cite{madry2017towards}. Other notable attacks include the Jacobian Saliency Map Adversary (JSMA), which leverages a saliency map derived from the Jacobian matrix to identify and perturb key pixels~\cite{papernot2016limitations}, and the Carlini and Wagner (C\&W) attack, an optimization-based approach designed to produce minimal, nearly invisible perturbations that still cause reliable misclassification~\cite{carlini2017towards}.

\citet{dong2020benchmarking} highlighted the importance of evaluating robustness using robustness curves that shows model performance across different level of perturbation and attack strengths which provides a more comprehensive assessment of adversarial robustness compared to traditional point-wise accuracy. The authors also discussed about the effectiveness of adversarial training and the robustness of randomization-based defenses against query-based black-box attacks. \citet{zhao2024evaluating} proposed comprehensive evaluation of adversarial robustness in VLMs. The results show that advanced models like  MiniGPT-4~\cite{zhu2023minigpt}, LLaVA~\cite{liu2024visual}, BLIP-2~\cite{li2023blip} trained on large datasets are vulnerable to adversarial noise, which results in generating erroneous or targeted output. By evaluating black-box attacks, the authors concluded that adversarial attacks crafted against one VLM can often be transferred to others, showing the widespread vulnerability across different architectures.

\section{Methodology}

Assessing the vulnerability of VLMs to adversarial attacks is essential for ensuring their reliability and safety in real-world applications. Existing metrics like CLEVER \cite{weng2018evaluating},
 AutoAttack \cite{croce2020reliable} often fail to capture the combined impact of random noise and targeted adversarial perturbations on model performance. Developing a comprehensive evaluation metric that integrates the effects of various perturbations can efficiently assess a model’s robustness
 
\begin{figure}[t]
\centering
\includegraphics[width=\columnwidth]{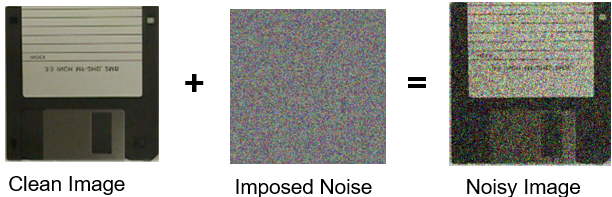} 
\caption{Imposing noise to clean image}
\label{noiseimage}
\end{figure}

\begin{figure*}[h]
\centering
\includegraphics[width=\textwidth]{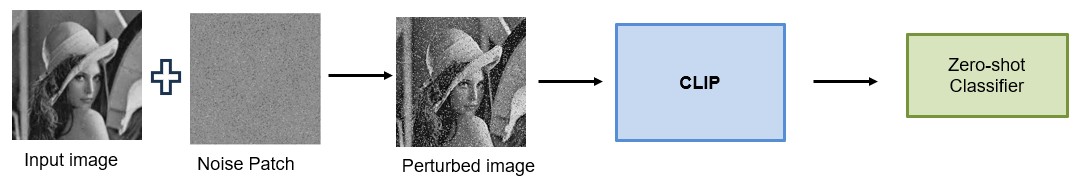} 
\caption{Methodology Process.}
\label{model}
\end{figure*}

  In our experiment, illustrated in Figure~\ref{model}, we conducted a classification task using the CLIP model, a vision-language
 model, on the Caltech-256 dataset. We randomly selected 300 images from the dataset, that is
 approximately 1\% of the entire dataset ensuring that all the classes were included. We applied
 each of the three previously mentioned noise types to our sample set. The implementation of the
 noises in as follows:
 \begin{itemize}
     \item  For each noise type, starting with a clean image, we increased the noise level by 0.01 in each iteration as shown in Figure \ref{noiseimage}. We can denote it as, for any noise $N(i)$ applied to an image at iteration i we add
 0.01 noise level at each iteration. $N(i + 1) = N(i) + 0.01$.
 \item At each noise level, we evaluated the CLIP model’s classification performance on the perturbed image.
 \item We tracked the specific noise level at which the model first misclassified the image, marking the threshold of the model's robustness for that particular image and noise type. Let $N_{final} = N(i)$ and suppose $F(I, N(i)) \neq True\_Label$, where $I$ is the original image and $F(I, N(i))$ is the model function for class prediction. In Table \ref{tab:gaussian-noise-sample}, we show a sample of noise levels and the corresponding model performance for Gaussian noise.
 \end{itemize}

 \begin{table}[t]
\centering
\begin{tabular}{r l l}
\toprule
\textbf{Noise Level} & \textbf{Original Class} & \textbf{Misclassified} \\ 
\midrule
0.75 & Telephone & Soda-Can \\
0.29 & Playing-Card & Sheet-music \\
0.10  & Airplane & Hotdog \\
0.46 & Saturn & Mussels \\
0.56 & Penguin & Chimp \\
\bottomrule
\end{tabular}
\caption{Sample of noise levels and model performance for Gaussian noise.}
\label{tab:gaussian-noise-sample}
\end{table}

\begin{figure}[h]
    \centering
    \begin{subfigure}[b]{0.2\textwidth}
        \centering
        \includegraphics[width=\textwidth]{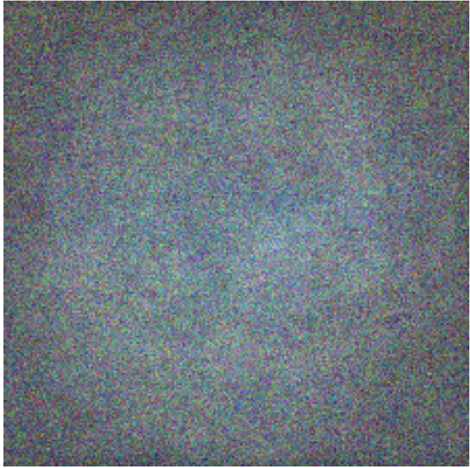}
        \caption{Gaussian Noise}
        \label{fig:image1}
    \end{subfigure}
    \hspace{1em}
    \begin{subfigure}[b]{0.2\textwidth}
        \centering
        \includegraphics[width=\textwidth]{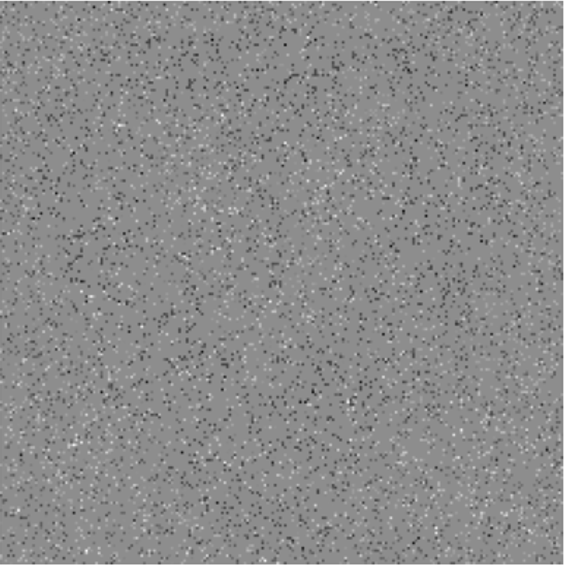}
        \caption{Salt and Pepper Noise}
        \label{fig:image2}
    \end{subfigure}
    \hspace{1em}
    \begin{subfigure}[b]{0.2\textwidth}
        \centering
        \includegraphics[width=\textwidth]{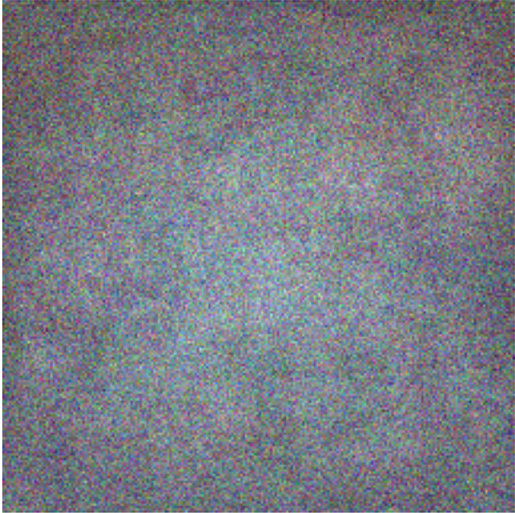}
        \caption{Uniform Noise}
        \label{fig:image3}
    \end{subfigure}

    \caption{Noise patches of three types of noises.}
    \label{fig:patches}
\end{figure}

 We gathered the noise levels that caused misclassification for each of the 300 images. By
 averaging these noise levels, we created a composite noise image patch representing the average
 noise pattern that leads to misclassification in the model. We performed this experiment for each
 of the three noises and collected the average noise patch as shown in Figure \ref{fig:patches}.

\section{Results and Performance Analysis}
Once we established both the noise patches and the saliency pattern, our next goal was to rigorously assess their potential as adversarial perturbations in comparison to a well-known benchmark, FGSM. We approached this comparison systematically. First, we introduced image-level perturbations derived from each of our identified noise patches. These patches, generated by averaging the noise levels at which misclassifications first occurred, were designed to capture the “average vulnerability” of the model to each type of noise. By applying these noise patches to previously unseen images, we aimed to determine how well they generalized as universal adversarial perturbation. In other words, we wanted to see if a single, pre-computed noise pattern could reliably induce misclassification in images it had never “seen” during the noise derivation phase.

We wanted to gain a deeper understanding of which pixels or regions of the image seems to be more vulnerable towards adversarial perturbation. Using the same noise patches, we generated a saliency pattern. This saliency pattern highlights image regions that, when altered, cause the model’s predictive confidence and accuracy to degrade. By applying this saliency pattern as a perturbation, we effectively tested the ability of focusing our adversarial attack on critical, high-impact areas within the image. This approach allowed us to move from broad-spectrum noise attacks to more targeted, strategic perturbations.

After experimenting with both the noise patches and the saliency pattern, we then introduced the conventional FGSM attack for comparison. While FGSM is known for its simplicity but as it calculates the gradients of loss to mislead the model, it is computationally expensive. By taking into account the result of FGSM, we could ascertain how well our noise patches and saliency-based perturbations measure up against a classic, widely recognized adversarial method. The outcomes of these experiments are summarized in Table~\ref{tab:attack-comparison}, which provides a direct comparison of the model’s accuracy for each type of attack. With the help of the results of Table \ref{tab:attack-comparison},  we developed a new metric that jointly considers the impact of random noise and targeted adversarial perturbations on the model’s performance. This metric offers a more comprehensive measure of the model’s vulnerability. Our evaluation metric is defined as:
\begin{table}[t!]
\centering
\begin{tabular}{l r}
\toprule
\textbf{Attack Name} & \textbf{Accuracy(\%)} \\
\midrule
Baseline & 95.00 \\
Gaussian Noise & 67.54 \\
Salt and Pepper & 66.80 \\
Uniform Noise & 66.56 \\
FGSM & 9.35 \\
\bottomrule
\end{tabular}
\caption{Comparison of Model Accuracy under Different Adversarial Attacks}
\label{tab:attack-comparison}
\end{table}
\[
\text{Noise Impact Score} = \frac{\text{Acc}_{\text{Baseline}} - \text{Acc}_{\text{Noise}}}{\text{Acc}_{\text{Baseline}}} \times 100 ,
\]
\[
\text{FGSM Impact Score} = \frac{\text{Acc}_{\text{Baseline}} - \text{Acc}_{\text{FGSM}}}{\text{Acc}_{\text{Baseline}}} \times 100 ,
\]
where $\text{Acc}_{\text{Baseline}}$ is the baseline accuracy, $\text{Acc}_{\text{Noise}}$ is the accuracy under random noise, and $\text{Acc}_{\text{FGSM}}$ is the accuracy under FGSM attack.

$\text{Vulnerability Score} = w_1 \times \text{Noise Impact Score} + w_2 \times \text{FGSM Impact Score}
$
where $w_1 + w_2 = 1$ and $w_1, w_2 \geq 0$.

By utilizing our proposed evaluation metric, we aim to systematically assess the vulnerability of VLMs to adversarial perturbations. This metric provides an efficient and scalable method for researchers to quantify their models' robustness, as demonstrated in our experiments where we worked with only 1\% of the dataset. Despite the limited sample size, the results effectively highlighted the sensitivity of the model to both random noise and targeted adversarial attacks, underscoring the metric's reliability.

A key advantage of our approach lies in its flexibility: the weights $w_1$ and $w_2$ can be adjusted to emphasize either random noise impact or adversarial perturbations, depending on the specific evaluation needs. This adaptability allows researchers to tailor the metric to suit different scenarios, such as prioritizing robustness against natural perturbations in real-world environments or focusing on resilience to deliberate adversarial attacks~\cite{rashid2024aisafety}.

\section{Conclusions}

In this work, we introduced a novel framework to evaluate the adversarial vulnerabilities of VLMs, focusing on public mission applications. Our method integrates the analysis of random noise and adversarial attacks, providing a comprehensive Vulnerability Score that highlights model weaknesses under diverse perturbation conditions. By leveraging incremental noise application and saliency-based analysis, we demonstrated the effectiveness of our approach in quantifying risks and identifying critical areas of vulnerability.

The proposed framework addresses a pressing need in public sector AI, ensuring that systems deployed in high-stakes environments maintain trust and reliability. While computational complexity poses a potential challenge, the insights gained from this evaluation provide actionable metrics for developers, policymakers, and organizations seeking to deploy robust AI systems. Future work will optimize the framework’s computational efficiency and extend its applicability to other multimodal AI architectures.

Our findings contribute to the broader effort of ensuring trustworthy AI in public service, offering a practical and scalable solution for assessing and improving the safety of VLMs in mission-critical applications.

\section{Acknowledgments}
Part of this work was funded by the National Science Foundation under grants CNS-2210091 and CNS-2136961, and by the U.S. Department of Education under grant P116Z230151.

\bibliography{aaai25}

\end{document}